\documentclass[twoside,twocolumn,9pt]{article}
\usepackage{extsizes}
\usepackage[super,sort&compress,comma]{natbib} 
\usepackage[version=3]{mhchem}
\usepackage[left=1.5cm, right=1.5cm, top=1.785cm, bottom=2.0cm]{geometry}
\usepackage{balance}
\usepackage{mathptmx}
\usepackage{sectsty}
\usepackage{graphicx} 
\usepackage{lastpage}
\usepackage[format=plain,justification=justified,singlelinecheck=false,font={stretch=1.125,small,sf},labelfont=bf,labelsep=space]{caption}
\usepackage{float}
\usepackage{fancyhdr}
\usepackage{fnpos}
\usepackage[english]{babel}
\addto{\captionsenglish}{%
  
}
\usepackage{array}
\usepackage{droidsans}
\usepackage{charter}
\usepackage[T1]{fontenc}
\usepackage[usenames,dvipsnames]{xcolor}
\usepackage{setspace}
\usepackage[compact]{titlesec}
\usepackage{hyperref}
\usepackage{siunitx}
\usepackage[switch]{lineno}

\definecolor{red}{rgb}{0,0,0}
\definecolor{blue}{rgb}{0,0,0.75}
\definecolor{green}{rgb}{0,0.5,0}
\newcommand{\red}[1]{{\color{red} #1}}

\usepackage{ulem}

\definecolor{cream}{RGB}{222,217,201}

\begin{document}

\pagestyle{fancy}
\thispagestyle{plain}
\fancypagestyle{plain}{
\renewcommand{\headrulewidth}{0pt}
}

\makeFNbottom
\makeatletter
\renewcommand\LARGE{\@setfontsize\LARGE{15pt}{17}}
\renewcommand\Large{\@setfontsize\Large{12pt}{14}}
\renewcommand\large{\@setfontsize\large{10pt}{12}}
\renewcommand\footnotesize{\@setfontsize\footnotesize{7pt}{10}}
\makeatother

\renewcommand{\thefootnote}{\fnsymbol{footnote}}
\renewcommand\footnoterule{\vspace*{1pt}%
\color{cream}\hrule width 3.5in height 0.4pt \color{black}\vspace*{5pt}} 
\setcounter{secnumdepth}{5}

\makeatletter 
\renewcommand\@biblabel[1]{#1}            
\renewcommand\@makefntext[1]%
{\noindent\makebox[0pt][r]{\@thefnmark\,}#1}
\makeatother 
\renewcommand{\figurename}{\small{Fig.}~}
\sectionfont{\sffamily\Large}
\subsectionfont{\normalsize}
\subsubsectionfont{\bf}
\setstretch{1.125} 
\setlength{\skip\footins}{0.8cm}
\setlength{\footnotesep}{0.25cm}
\setlength{\jot}{10pt}
\titlespacing*{\section}{0pt}{4pt}{4pt}
\titlespacing*{\subsection}{0pt}{15pt}{1pt}

\fancyfoot{}
\fancyfoot[LO,RE]{\vspace{-7.1pt}\includegraphics[height=9pt]{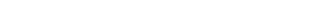}}
\fancyfoot[CO]{\vspace{-7.1pt}\hspace{13.2cm}\includegraphics{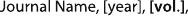}}
\fancyfoot[CE]{\vspace{-7.2pt}\hspace{-14.2cm}\includegraphics{head_foot/RF}}
\fancyfoot[RO]{\footnotesize{\sffamily{1--\pageref{LastPage} ~\textbar  \hspace{2pt}\thepage}}}
\fancyfoot[LE]{\footnotesize{\sffamily{\thepage~\textbar\hspace{3.45cm} 1--\pageref{LastPage}}}}
\fancyhead{}
\renewcommand{\headrulewidth}{0pt} 
\renewcommand{\footrulewidth}{0pt}
\setlength{\arrayrulewidth}{1pt}
\setlength{\columnsep}{6.5mm}
\setlength\bibsep{1pt}

\makeatletter 
\newlength{\figrulesep} 
\setlength{\figrulesep}{0.5\textfloatsep} 

\newcommand{\topfigrule}{\vspace*{-1pt}%
\noindent{\color{cream}\rule[-\figrulesep]{\columnwidth}{1.5pt}} }

\newcommand{\botfigrule}{\vspace*{-2pt}%
\noindent{\color{cream}\rule[\figrulesep]{\columnwidth}{1.5pt}} }

\newcommand{\dblfigrule}{\vspace*{-1pt}%
\noindent{\color{cream}\rule[-\figrulesep]{\textwidth}{1.5pt}} }

\makeatother

\twocolumn[
  \begin{@twocolumnfalse}
{\includegraphics[height=30pt]{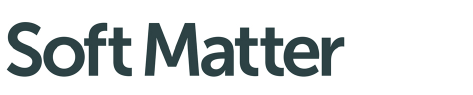}\hfill\raisebox{0pt}[0pt][0pt]{\includegraphics[height=55pt]{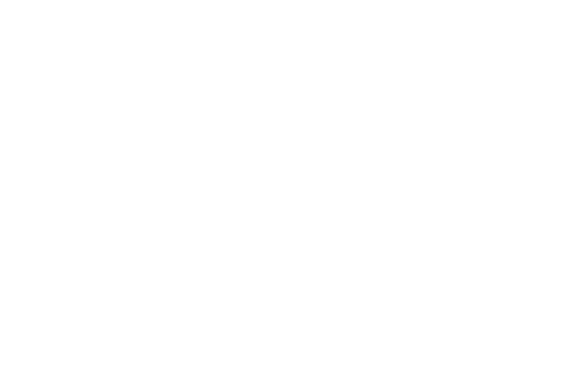}}\\[1ex]
\includegraphics[width=18.5cm]{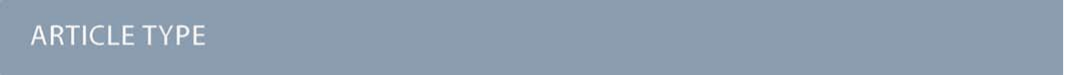}}\par
\vspace{1em}
\sffamily
\begin{tabular}{m{4.5cm} p{13.5cm} }

\includegraphics{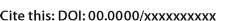} & \noindent\LARGE{\textbf{Topological defects in multi-layered swarming bacteria}} \\
\vspace{0.3cm} & \vspace{0.3cm} \\

 & \noindent\large{Victor Yashunsky,\textit{$^{a}$} Daniel J.G. Pearce,\textit{$^{b}$} Gil Ariel,\textit{$^{c}$} Avraham Be’er\textit{$^{d, e}$}} \\

\includegraphics{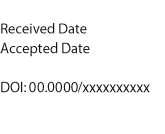} & \noindent\normalsize{Topological defects, which are singular points in a director field, play a major role in shaping active systems. Here, we experimentally study topological defects and the flow patterns around them, that are formed during the highly rapid dynamics of swarming bacteria. The results are compared to the predictions of two-dimensional active nematics. We show that, even though some of the assumptions underlying the theory do not hold, the swarm dynamics is in agreement with two-dimensional nematic theory. In particular, we look into the multi-layered structure of the swarm, which is an important feature of real, natural colonies, and find a strong coupling between layers. Our results suggest that the defect-charge density is hyperuniform, i.e., that long range density-fluctuations are suppressed.
}
\end{tabular}

 \end{@twocolumnfalse} \vspace{0.6cm}

  ]

\renewcommand*\rmdefault{bch}\normalfont\upshape
\rmfamily
\section*{}
\vspace{-1cm}


\footnotetext{\textit{$^{a}$~The Swiss Institute for Dryland Environmental and Energy Research, The Jacob Blaustein Institutes for Desert Research, Ben-Gurion University of the Negev, Sede Boqer Campus, 84990 Midreshet Ben‑Gurion, Israel; E-mail: yashunsk@bgu.ac.il;}}
\footnotetext{\textit{$^{b}$~Department of Theoretical Physics, University of Geneva, 1211 Geneva, Switzerland;}}
\footnotetext{\textit{$^{c}$~Department of Mathematics, Bar-Ilan University, 52900 Ramat‑Gan, Israel; E-mail: arielg@math.biu.ac.il;}}
\footnotetext{\textit{$^{d}$~Zuckerberg Institute for Water Research, The Jacob Blaustein Institutes for Desert Research, Ben-Gurion University of the Negev, Sede Boqer Campus, 84990 Midreshet Ben‑Gurion, Israel;}}
\footnotetext{\textit{$^{e}$~The Department of Physics, Ben-Gurion University of the Negev, 84105 Beer‑Sheva, Israel; E-mail: beera@bgu.ac.il.}}

\footnotetext{\dag~Electronic Supplementary Information (ESI) available: DOI: XXX/xxx/}


\section{Introduction}
Active matter has become a central topic of contemporary physics, with examples ranging from intra-cellular processes to animal herds \cite{RefWorks:RefID:33-bowick2022symmetry, RefWorks:RefID:48-shankar2022topological}. Active particles are typically classified as either polar or nematic, depending on their underlying symmetries or interactions in the system \cite{RefWorks:RefID:36-doostmohammadi2018active}. In the nematic case, flows are driven by active stresses. It has been shown that above a certain level of activity, nematic systems develop chaotic collective swirling motion, commonly addressed as “active nematic turbulence” \cite{RefWorks:RefID:26-alert2020universal, RefWorks:RefID:36-doostmohammadi2018active}. This highly active nematic regime is manifested in the spontaneous generation and annihilation of $+1/2$ and $-1/2$ topological defect pairs in the director field, and was first demonstrated in suspensions of microtubule bundles \cite{RefWorks:RefID:47-sanchez2012spontaneous}. 
Modeling and numerical simulations explored the properties of defects and provide a comprehensive understanding of their dynamics \cite{pismen2013dynamics,giomi2013defect,thampi2014instabilities,RefWorks:RefID:38-giomi2015geometry,putzig2016instabilities,tang2017orientation,cortese2018pair,shankar2018defect,shankar2019hydrodynamics,pearce2021orientational,kozhukhov2022mesoscopic,RefWorks:RefID:33-bowick2022symmetry,palmer2022understanding,RefWorks:RefID:46-pearce2021properties,RefWorks:RefID:61-vromans2016orientational}.
Giomi \cite{RefWorks:RefID:38-giomi2015geometry} predicted that an active nematic turbulence phase should have an intrinsic length that determines the average vortex size, set by the balance between elastic and active forces. This prediction was later verified in active nematic films of microtubules \cite{RefWorks:RefID:43-keber2014topology}, epithelial cells \cite{RefWorks:RefID:32-blanch-mercader2018turbulent}, and cytoskeletal reconstitutions \cite{RefWorks:RefID:39-guillamat2017taming}. An additional intriguing result was observed in suspensions of microtubule bundles, revealing that $+1/2$ defects may exhibit orientational ordering on scales significantly exceeding the mean free path and the lifetime of individual defects \cite{RefWorks:RefID:35-decamp2015orientational,putzig2016instabilities,shankar2018defect}. The occurrence of long-range order has been challenged in \cite{pearce2021orientational}.

\begin{figure}[!h]
\centering
  \includegraphics[width=0.5\textwidth]{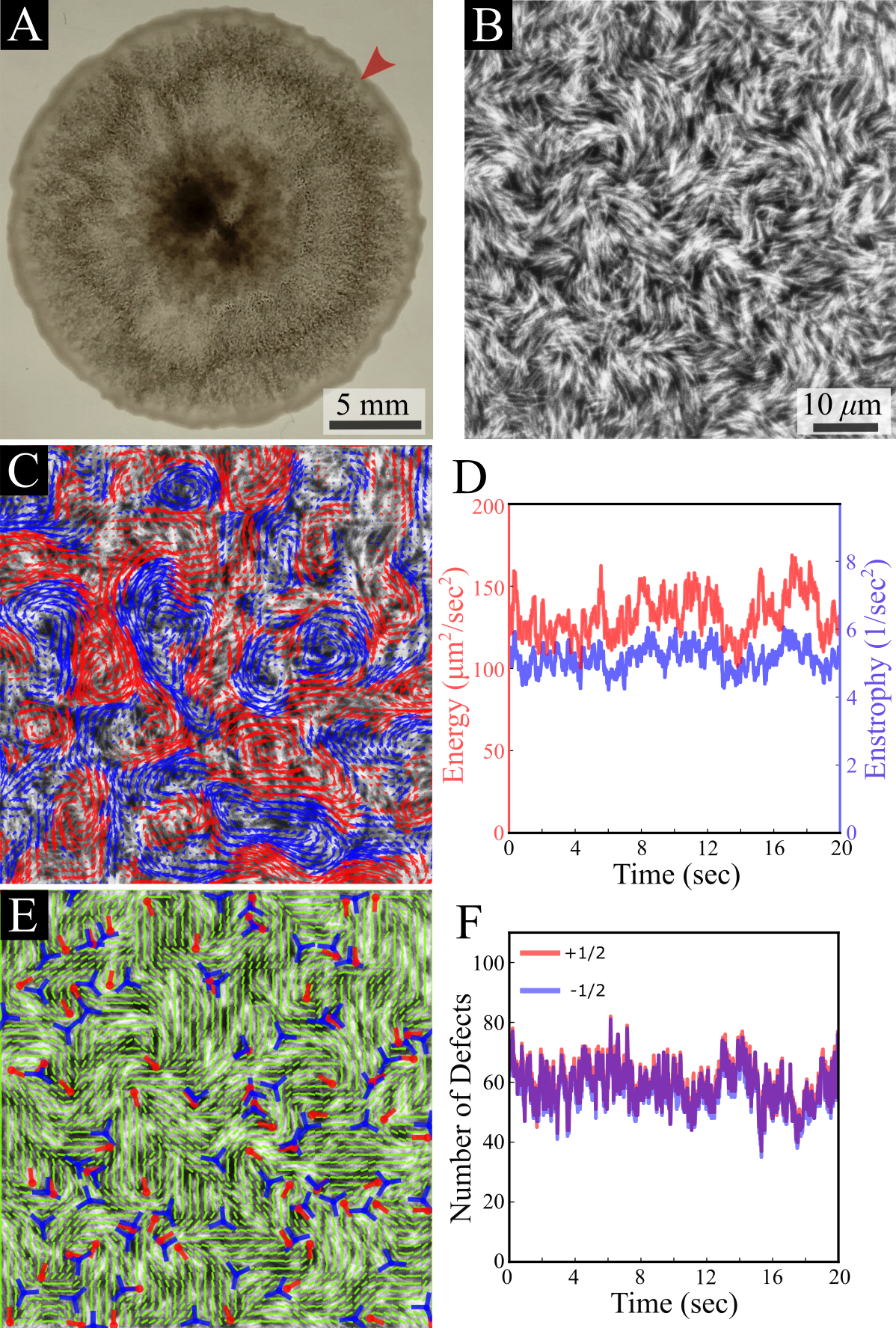}
  \caption{\textit{B. subtilis} swarming. (A) A top-view macroscopic image of a \textit{B. subtilis} colony; the arrow indicates the region of interest shown in B. (B) A top-view microscopic image of fluorescently labeled \textit{B. subtilis} swarming cells taken with a $40\times$ objective. The field of view is \qtyproduct{150 x 150}{\um}. (C) Velocity field (in the background of the cells) calculated using an optical flow method, demonstrating chaotic collective swirling motion. Blue and red colors indicate clockwise and counterclockwise rotating regions. (D) Time evolution of enstrophy $\Omega$ (blue) and kinetic energy per unit mass $E$ (red). (E) The nematic director field extracted from (B) with positive (red) and negative (blue) locations and orientations of half-integer defects.  (F) Time evolution of the observed number of $+1/2$ (red) and $-1/2$ (blue) defects.}
  \label{Fig.1}
\end{figure}

Topological defects were studied in a variety of rod-shaped motile bacterial systems as well. Slowly gliding \textit{Myxococcus xanthus} were found to generate defects that promote layer formation in three-dimensional colonies \cite{RefWorks:RefID:34-copenhagen2021topological}. Studies of \textit{Bacillus subtilis} biofilms, which are also slowly moving systems, have shown that a living nematic can actively shape itself and its boundary to regulate its internal architecture through growth-induced stresses \cite{RefWorks:RefID:45-nijjer2023biofilms}, and that localized stress and friction drive buckling and edge instabilities which further create nematically aligned structures and topological defects \cite{RefWorks:RefID:52-yaman2019emergence, RefWorks:RefID:56-shimaya2022tilt-induced}. Growing single-layer micro-colonies of \textit{Escherichia coli} develop topological defect pairs that migrate towards the periphery \cite{dell2018growing}. \textit{B. subtilis} cells rapidly swimming in liquid crystal suspensions have demonstrated the role of topological defects in the distribution and accumulation of cells in the crowd \cite{peng2016command,RefWorks:RefID:37-genkin2017topological}. 

Swarming bacteria, which are self-propelled elongated cells, grow on surfaces and form highly dynamic swirling patterns. Li et al. \cite{RefWorks:RefID:44-li2019data-driven} have studied topological defects in a thin (2-3 layers) flow of swarming \textit{Serratia marcescens}, and found $+1/2$ and $-1/2$ defects. Their results were surprising because swarming bacteria are propelled forward against the direction of the flagella, making them active polar particles. However, the principle interactions between cells are steric and hydrodynamic, approximately a dipole \cite{RefWorks:RefID:30-ariel2018collective}, which are both nematic. 

In this paper, we describe a comprehensive analysis of defects in swarming colonies of \textit{B. subtilis}. First, we consider the colony as a two dimensional (2D) system. Because the colony is flat and uniform, a quasi-2D approximation has been prevalent in most experimental and theoretical studies of bacterial systems \cite{RefWorks:RefID:31-be’er2019statistical,RefWorks:RefID:42-jeckel2019learning,RefWorks:RefID:58-wensink2012meso-scale,RefWorks:RefID:29-aranson2022bacterial, RefWorks:RefID:69-meacock2021bacteria,RefWorks:RefID:70-dunkel2013fluid}. The experimental results are compared to the predictions of active nematic theory. We demonstrate that, as far as defects are concerned, a single layer of swarming bacteria may indeed be considered as an active nematic system. However, a naturally growing colony is not two dimensional, but has a non-negligible thickness, here approximately the width of $7$ cells. To this end, we simultaneously capture the swarm dynamics of two separated planar fields of view at different depths and compare their dynamics. We conclude with a discussion of the main assumptions underlying the theory of 2D active nematics, the extent to which they hold in bacterial swarms, and the implications on the physics of swarming bacteria.

\section{Results}
\subsection{Single-layer analysis}
Our results are carried out using \textit{B. subtilis} swarm cells \qtyproduct{1 x 7}{\um} that grow on semi-solid agar surfaces. The colony of cells forms a flat active structure with a uniform thickness ($\sim$\SI{7}{\um}; thickness of $7$ cells) across centimeter-scale distances (Fig. \ref{Fig.1}A and MOV. S1). Observations are done on a \qtyproduct{150 x 150}{\um} window ($40\times$ magnification), enabling precise tracking of the cell flow and reliable statistics with a large number of cells. Top-view images of the colony (Fig. \ref{Fig.1}B) show that cells form swirling patterns with vortices rotating in clockwise and counterclockwise directions (Fig. \ref{Fig.1}C). For the remainder of this section, we describe results for fluorescently labeled cells in a layer close to the bottom of the colony. The energy \red{$E=\langle(v_x^2+v_y^2)/2\rangle$} and enstrophy \red{$\Omega=\langle\omega^2/2\rangle$, where $\langle \cdot \rangle$ signifies spatial average}, obtained from the velocity \red{$\textbf{\textit{v}}=(v_x, ~v_y)$} and vorticity \red{$\omega=\partial_xv_y-\partial_yv_x$} fields, are largely constant in time (Fig. \ref{Fig.1}D), indicating that the system is in a steady-state on the experimentally relevant time scale. The director field can be extracted using structure factor methods and features half-integer ($\pm$1/2) nematic defects (Fig. \ref{Fig.1}E and MOV. S1). The number of all defects fluctuates with time, but keeps a constant average per unit area for the observation time (about $20$ sec), with a similar value of positive and negative ones, indicating a close to zero total charge at all times (Fig. \ref{Fig.1}F). These results indicate that while the system is in motion and far from equilibrium, it has reached a steady state similar to the aforementioned ``active turbulence'' \cite{RefWorks:RefID:26-alert2020universal, RefWorks:RefID:36-doostmohammadi2018active}.

\begin{figure*}
\centering
  \includegraphics[width=0.75\textwidth]{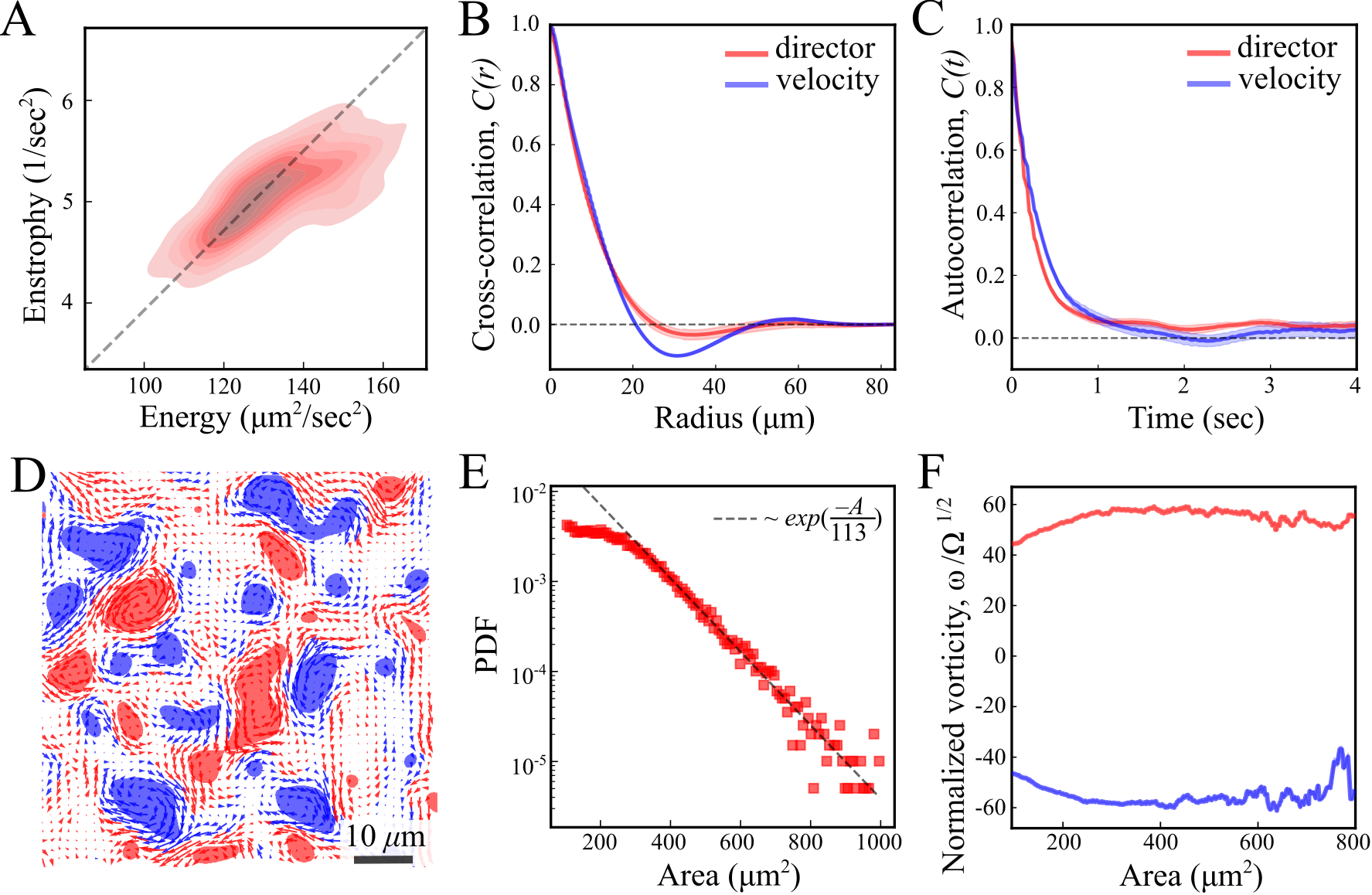}
  \caption{Statistical analysis of the director and velocity fields. (A) Enstrophy $\Omega$ vs. kinetic energy $E$ for each timepoint throughout the entire experiment, presented as a density map (B, C) Cross-correlation and autocorrelation functions of the director (red) and velocity (blue) fields. 
  (D) Okubo-Weiss field thresholded to negative values, $Q<0$. Blue and red domains denote clockwise and counterclockwise rotating regions. (E) Log-linear representation of the vortex area probability density distribution. The dotted line shows an exponential fit. (F) The mean vorticity per vortex is normalized by the square root of enstrophy $\omega / \Omega^{1/2}$ as a function of the vortex area. Blue and red curves represent clockwise and counterclockwise rotating vortices.}
  \label{Fig.2}
\end{figure*}

As predicted for turbulent active nematic systems, energy and enstrophy are correlated (Fig. \ref{Fig.2}A). The ratio between the two introduces an intrinsic length scale (Fig. \ref{Fig.2}A)\cite{RefWorks:RefID:38-giomi2015geometry,RefWorks:RefID:70-dunkel2013fluid}. Its value, $l^2=4E/\Omega=104 \pm 10$ \SI{}{\square\um} (mean $\pm$ standard deviation) agrees with the characteristic decay length of spatial correlations \red{$l=\langle \textbf{f}(x,y) \cdot \textbf{f}(x+x',y+y')\rangle_{(x,y)}$} of the director and velocity fields (Fig. \ref{Fig.2}B), $l_{\rm d}= 10.41\pm$\SI{0.03}{\um} and $l_{\rm v}=11.28 \pm$\SI{0.03}{\um}, respectively. Temporal correlation \red{$\tau=\langle \textbf{f}(t) \cdot \textbf{f}(t+t')\rangle_t$} times from the director and velocity fields (Fig. \ref{Fig.2}C) introduce a time scale of $\tau_{\rm d} = 0.25\pm 0.01$ sec and $\tau_{\rm v} = 0.30 \pm 0.01$ sec (mean $\pm 95$ \% confidence interval), respectively. These results are compatible with previous experimental results in multilayer swarms \cite{RefWorks:RefID:41-ilkanaiv2017effect,RefWorks:RefID:31-be’er2019statistical,RefWorks:RefID:42-jeckel2019learning,RefWorks:RefID:44-li2019data-driven}. We analyse the flow by first calculating the Okubo-Weiss criterion $Q=-det \nabla \textbf{\textit{v}}<0$, which segments vortex areas by identifying coherent elliptic structures in the 2D flow field \cite{RefWorks:RefID:62-benzi1988self-similar} (Fig. \ref{Fig.2}D). An analysis of the vortex areas unveils an approximately exponential distribution (Fig. \ref{Fig.2}E), with a characteristic vortex area of $113\pm$\SI{5}{\square\um} (fit $\pm$ 95 \% confidence interval). Thus, all length scales are consistent with the prediction of a single intrinsic length scale\footnote{The length-scale value derived from the enstrophy and energy ratio strongly depends on the spatial filter applied to the flow field because such statistics depend on derivatives in the flow field. This is in contrast to the Okubo-Weiss defect segmentation, which does not depend much on the smoothing.}. Additionally, we observe that the mean vorticity per vortex, $\omega / \Omega^{1/2}$, remains approximately size-independent with respect to the vortex area (Fig. \ref{Fig.2}F). We find no signs of a break in chiral symmetry, in contrast to some other cellular active nematics\cite{RefWorks:RefID:53-yashunsky2022chiral}. 

\begin{figure}
\centering
  \includegraphics[width=0.49\textwidth]{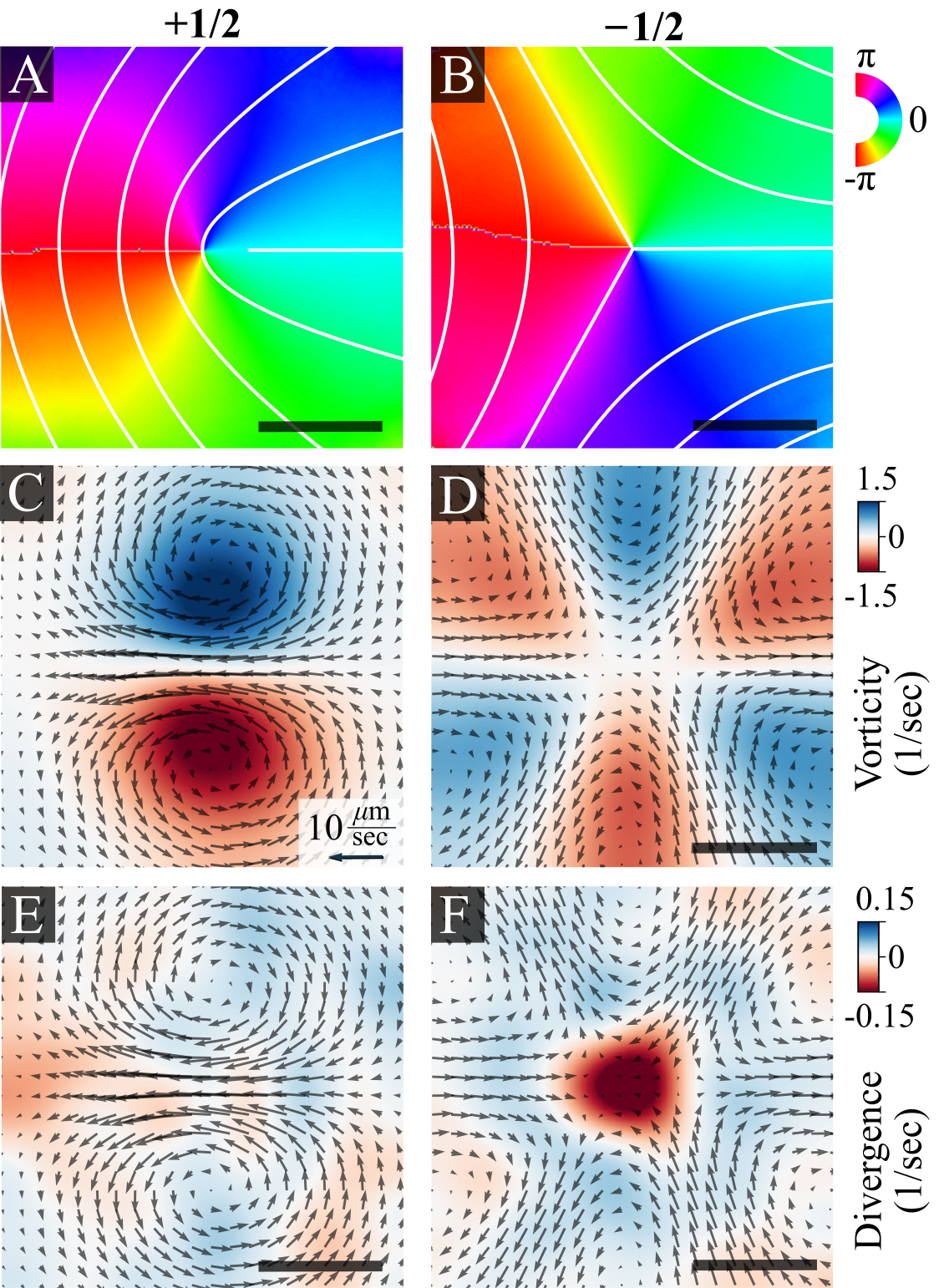}
  \caption{\red{Ensemble average of director and flow fields near $+1/2$ and $-1/2$ nematic defects, as obtained from experiments, centered and aligned. (A, B) Director fields depicting $+1/2$ and $-1/2$ defects. The color maps and overlaid white lines show the ensemble average of the director orientation, varying from $-\pi$ to $\pi$.
  (C-D) Average velocity fields. Colors indicate the vorticity (C, D) and divergence (E, F) for $+1/2$ (left) and $-1/2$ (right) defects. The superimposed black arrows indicate the direction of the velocity field. The magnitude of the velocity field is indicated by the black arrow, corresponding to a value of \SI{10}{\um}/sec}. The scale bar is \SI{10}{\um}.}
  \label{Fig.3}
\end{figure}

Figure \ref{Fig.3} shows the average bacterial orientation and velocity around $+1/2$ and $-1/2$ defects. See the Methods section for details. We find a symmetric structure of vortex pairs in agreement with active nematic hydrodynamics\cite{giomi2013defect}. The cores of $+1/2$ defects are surrounded by a pair of counter-rotating vortices (Fig. \ref{Fig.3}C, E). Consequently, there is a net flow at the core of a $+1/2$ defect and it self-propels with an average speed of $10 \pm 2$ \SI{}{\um}/sec (mean $\pm$ standard deviation)\red{, while the average speed of the swarm is $16 \pm 5$ \SI{}{\um}/sec}. The $+1/2$ defects self propel away from the tail (where the nematic director is oriented radially from the core of the defect). This implies that the cells exert an extensile active stress, in which bacteria push out along their long axis. The cores of $-1/2$ defects are encircled by three pairs of counter-rotating vortices that align with the defect's three-fold symmetry (Fig. \ref{Fig.3}D, F). The non-zero divergence of the velocity field indicates compressible flow, which is consistent with observations on bacterial active nematics \cite{you2018geometry,you2019mono,you2021confinement}. However, we observe a distinct influx toward the core of the $-1/2$ defects (Fig. \ref{Fig.3}F). \red{Due to the short defect lifetimes, of the order of 1 second, the actual degree of divergent flow is small. At the center of a defect this would correspond to a total compression of about 10 \% of a cell. Physically, this can be accommodated by compression of the flagellar cloud around each cell, which is compressible.} This \red{is different} from prior experimental findings with active microtubules \cite{RefWorks:RefID:47-sanchez2012spontaneous,RefWorks:RefID:43-keber2014topology}, mammalian cells \cite{RefWorks:RefID:32-blanch-mercader2018turbulent, RefWorks:RefID:59-kawaguchi2017topological, RefWorks:RefID:50-saw2017topological, RefWorks:RefID:49-sarkar2023crisscross}, slow moving bacteria \cite{RefWorks:RefID:34-copenhagen2021topological,RefWorks:RefID:45-nijjer2023biofilms,RefWorks:RefID:52-yaman2019emergence,RefWorks:RefID:56-shimaya2022tilt-induced} and highly active swimming bacteria \cite{RefWorks:RefID:37-genkin2017topological,genkin2018spontaneous}, which all reported accumulation at $+1/2$ defects but depletion close to $-1/2$ ones. \red{See the discussion section for the physical implications of this result.} 

\begin{figure*}[h!]
\centering
  \includegraphics[width=0.8\textwidth]{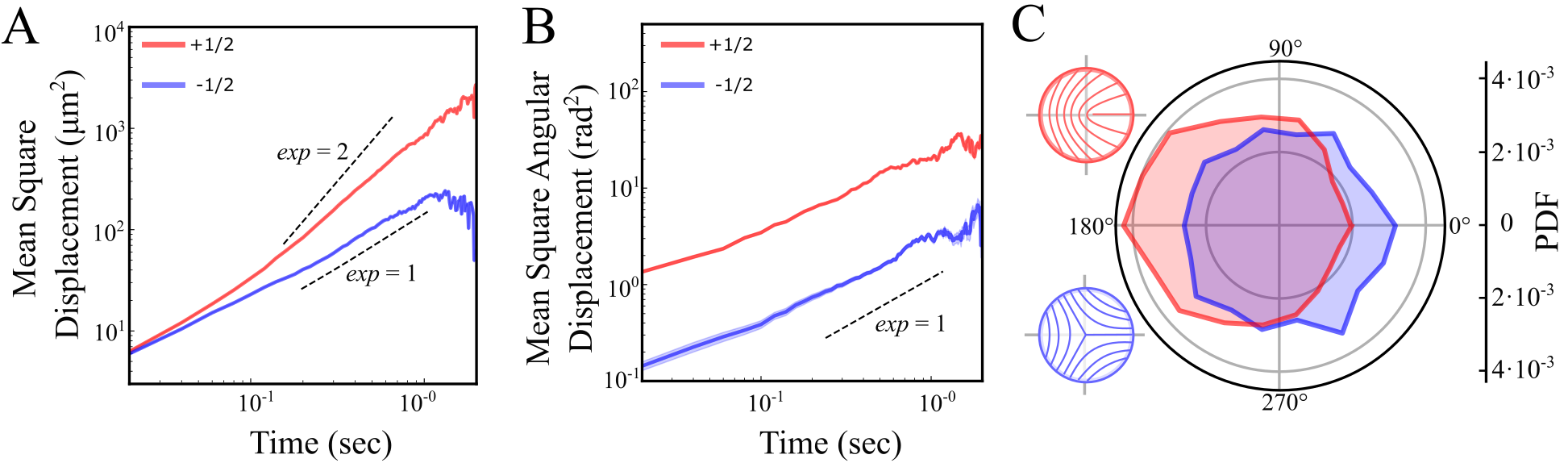}
  \caption{Statistical analysis of $+1/2$ and $-1/2$ defect trajectories. (A, B) Averaged mean square displacement (MSD) of defect position (A) and direction (B) for $+1/2$ (red) and $-1/2$ (blue) defects on a log-log scale. The dashed lines are guides to the eye with slopes 1 (diffusive) and 2 (ballistic). (C) Probability density distribution of the relative angle between the velocity vector and the defect orientation. Sketches illustrate the nematic orientation around $+1/2$ (red) and $-1/2$ (blue) defects.}
  \label{Fig.4}
\end{figure*}

Next, we examine defect trajectories. The mean squared displacement (MSD) of $+1/2$ defects adheres to a power law with an exponent of close to $2$, \red{signifying super-diffusion, possibly ballistic forward movement (Fig. \ref{Fig.4}A, C), at least on a time scale of $0.3 - 1$ seconds}. On shorter time scales, the nature of the dynamics is unclear due to the inherent noise in identifying defects. On longer time scales, trajectories were constrained by both the field of view and annihilation events, and therefore seldom extended beyond $1$ second. Conversely, $-1/2$ defects demonstrate diffusive behavior without any noticeable preferential direction (Fig. \ref{Fig.4}A, C). This is in contrast with preditions which show diffusive behaviour for both $+1/2$ and $-1/2$ defects at all but very short timescales \cite{RefWorks:RefID:38-giomi2015geometry}. Rotational diffusion is observed for both defect types (Fig. \ref{Fig.4}B). Hence, at long time scales, unless annihilating, we anticipate that the rotation of defect orientations will lead to a diffusive motion for $+1/2$ defects as well. The diffusive nature of the motion of the defects suggest that the elastic interactions between defects are very small as these can lead to long time correlations in the translation and rotation of defects \cite{RefWorks:RefID:46-pearce2021properties}.

The interaction between defects can be partially described by the structure factor, which characterizes the spatial correlations of density fluctuations as a function of the wave vector (see Methods for definitions). Figure \ref{Fig.5} shows the log of two structure factors. First, $S_{\rm both}(q)$ (red line) is the structure factor of both $+1/2$ and $-1/2$ defects, ignoring the sign, as a function of the norm of the wave vector, $q=|{\bf q}|$. See the methods section for details. For small $q$ values, $S_{\rm both}(q)>1$, indicating long-range attraction. Taking the defect sign into account is obtained by considering the charge-density structure factor, $S_\rho(q)$ \cite{RefWorks:RefID:40-hansen2013theory}. See the methods section for the precise definition. Our results (Fig.  \ref{Fig.5}, blue line) suggest that the charge-density structure factor vanishes in the limit of vanishing 
wave vector. This is a hallmark of hyperuniformity and is consistent with the local depletion in similar charges observed in microtubule based active nematics \cite{pearce2021orientational}. Specifically, plotted on a log-log scale, we find that approximately $S_\rho(q)= q^2$. Note that the results are limited by the minimal wave vector accessible due to the system size, $q_{\rm min}= 2\pi/L$, $L$ being the length of the observation frame, which is \SI{150}{\um} here.

\subsection{Multilayer analysis}
The colony thickness is taken into account using a splitting device (see the Methods section) that enables simultaneous tracking of the top and bottom layers of the colony. We use a 1:1 mix of two fluorescently labeled strains, on each we focus on a different height (green bottom and red top). We analyzed two layers that are \SI{7}{\um} apart, corresponding to 7 times the single-cell thickness. Despite the separation between layers, strong alignment is observed between their director and velocity fields (\ref{Fig.6}A, B). Quantitatively, the distribution of the phase between the vectors in the bottom ($h1$) and the top ($h2$) layers is quite narrow, with a coefficient of variation (the ratio between standard deviation and a half of the distribution range) of 18.1 \% and 12.1 \% for the director and flow phase distribution, respectively (Fig. \ref{Fig.6}C, D). 

Focusing on the defects, we first observe that analyzed independently, the defect statistics are very similar (Fig. S2). The defects are highly correlated, both in their location and direction (Fig.  \ref{Fig.7}), this is consistent with our observation that the nematic textures are highly correlated. In order to quantify this observation, we assign each defect its closest pair with the same sign from the other layer (Fig.  \ref{Fig.7}A). See Methods for details. Figure \ref{Fig.7}B depicts the probability to find a defect pair separated by a given distance (projected onto the plane), normalized by the same probability obtained for random uniform distributions. In other words, a value larger (lower) than 1 indicates that the probability to find the pair is higher (lower) than random. The results indicate a strong attraction between defects of the same charge across the layers. The angle of the relative position is uniform (Fig. \ref{Fig.7}B inset). This would indicate that the point like defects we observe on the top and bottom of the colony are connected by a line defect running through the bulk. The energy of this line defect increases with its length, hence it is minimized when defects are aligned directly above one another.  

Figure~\ref{Fig.7}C shows the phase distribution between the corresponding defects, $\varphi^{\pm 1/2}=\theta_{h1}^{\pm 1/2}-\theta_{h2}^{\pm 1/2}$, where $\theta_{h}^{\pm 1/2}$ is the direction (relative to the $x$-axis) of the $+1/2$ or $-1/2$ defect in layer \textit{h}, as illustrated in Fig. \ref{Fig.7}. The difference is taken modulus of the symmetry. \red{On average, defects tend to align with their corresponding defect in the other layer, as evidenced by the peak at zero in the distribution of phase differences (Fig. \ref{Fig.7}C).} The coefficient of variation of the defect angle phase (24.8 \% and 31.4 \%) is notably higher than the coefficient of variation of the phase between director fields (Fig. \ref{Fig.7}C). \red{This is likely due to the fact that the defects being compared often have a greater separation than just the distance between the layers.}

Figure~\ref{Fig.7}D demonstrates that for both defect types, the standard deviation of the phase \red{difference} increases proportionally with distance, indicating that the probability of misalignment grows with distance. \red{There are two likely explanations for this. First, as the distance between two defects increases, the nematic between them becomes increasingly unstable to distortions caused by the active stress, decreasing their correlation. Second, due to self propulsion initially misaligned $+1/2$ defects will move in different directions, increasing their separation over time relative to aligned defects~\cite{shankar2018defect}.} The direction becomes independent at a distance of about \SI{12}{\um} (Fig. \ref{Fig.7}D), marking the range of interaction between defects in different layers. This range is comparable to the characteristic size of vortices (Fig. \ref{Fig.2}E). This could be due to a de-correlation of the two ends of the line defect, or a mismatching of the defects between the two layers, as the vortex size is also comparable to the inter-defect distance.

\red{The small observed phase difference between the two layers of the colony in either the director field, Fig.~\ref{Fig.6}C, or the defect orientations, Fig.~\ref{Fig.7}C, indicates there is very little twist deformation in the bulk of colony.}

\section{Discussion}
The comparison between the two dimensional dynamics of the top (or bottom) surface of the swarm and the theory of 2D active nematics demonstrated very good agreement. Notably, we established the emergence of a single length scale and a well defined constant defect density. The areas of the vortices in the flow follow an exponential distribution with a consistent mean value (Fig.~\ref{Fig.2}E), and the vortex rotational frequency is approximately size-independent (Fig.~\ref{Fig.2}F). These two features serve as a distinctive signature of turbulence in active nematics \cite{RefWorks:RefID:38-giomi2015geometry}. This presents strong evidence that the behaviour of a multilayered colony of motile bacteria is indeed described by the physics of a two dimensional active nematic. 

The agreement between the bacterial colony and an active nematic is quite surprising given the differences between the two systems. For example, the bacteria dissipate energy and momentum into the agar substrate \cite{RefWorks:RefID:31-be’er2019statistical}. It is also well known that bacteria release surfactants and pump water from the agar in order to create the hydration layer in which they move \cite{RefWorks:RefID:31-be’er2019statistical}. Collisions between bacteria have been shown to likely be viscoelastic in nature, which can lead to stress aligning behaviour~\cite{you2021confinement}. As a result, the elastic constants and active stress likely depend on the pressure and local alignment of the bacteria \cite{you2018geometry}.

On the microscopic level, particles (here \textit{B. subtilis} cells) are essentially polar \cite{RefWorks:RefID:58-wensink2012meso-scale,RefWorks:RefID:30-ariel2018collective,RefWorks:RefID:42-jeckel2019learning,RefWorks:RefID:29-aranson2022bacterial}. They have a well defined tail marked by the flagellar bundle. As a result, cells generate a flow that is not exactly symmetric around their axis \cite{RefWorks:RefID:64-spagnolie2012hydrodynamics}. Still, it has been argued that the main interactions between cells are steric repulsion and dipole-like hydrodynamic interactions \cite{RefWorks:RefID:64-spagnolie2012hydrodynamics,RefWorks:RefID:30-ariel2018collective}, both of which are nematic. Our results are consistent with these assumptions.

Finally, the bacterial colony is multilayered with a thickness of about \SI{7}{\um}, about the length of a cell. Due to imaging constraints, we are not able to probe the full three dimensional structure of the colony but instead, image the two dimensional behaviour of the top or bottom surface. Our results show strong correlations between the orientation and velocity fields across the width of the system (Fig.~\ref{Fig.6}). \red{It has been theoretically shown that when the thickness of an active nematic liquid crystal increases, it becomes unstable to twist deformations in the bulk~\cite{shendruk2018twist}. Thus, our results indicate that the threshold thickness for this instability is greater than $7\mu m$ in bacterial colonies.} This supports our conclusion that a thin film approximation of the colony is applicable and the system can be well described as a two dimensional material. Furthermore, the location and orientation of topological defects on the two surfaces are highly correlated (Fig.~\ref{Fig.7}B, C). These are in fact the end points of a line defect that runs from the top to the bottom surface.

\begin{figure}[!h]
\centering
  \includegraphics[width=0.25\textwidth]{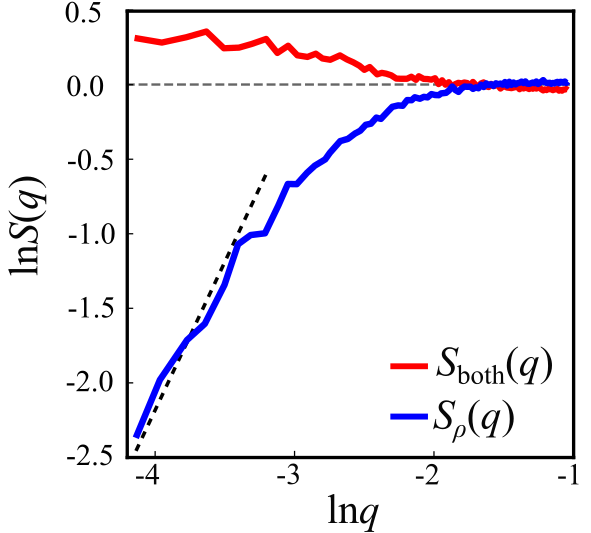}
  \caption{Defect structure factors. Red curve: the structure factor of the overall defect density, $S_{\rm both}(q)$, shows long range attraction. Blue curve: the structure factor for the defect charge, $S_\rho(q)$, shows reduction at small $q$ values, suggesting suppression of large-scale fluctuations. The dashed line is a guide to the eye with slope $2$, the theoretical prediction for ionic solutions at equilibrium. The vanishing of $S_\rho(q)$ in the limit $q\xrightarrow ~0$ is the hallmark of hyperuniformity.
}
  \label{Fig.5}
\end{figure}

\begin{figure}
\centering
  \includegraphics[width=0.5\textwidth]{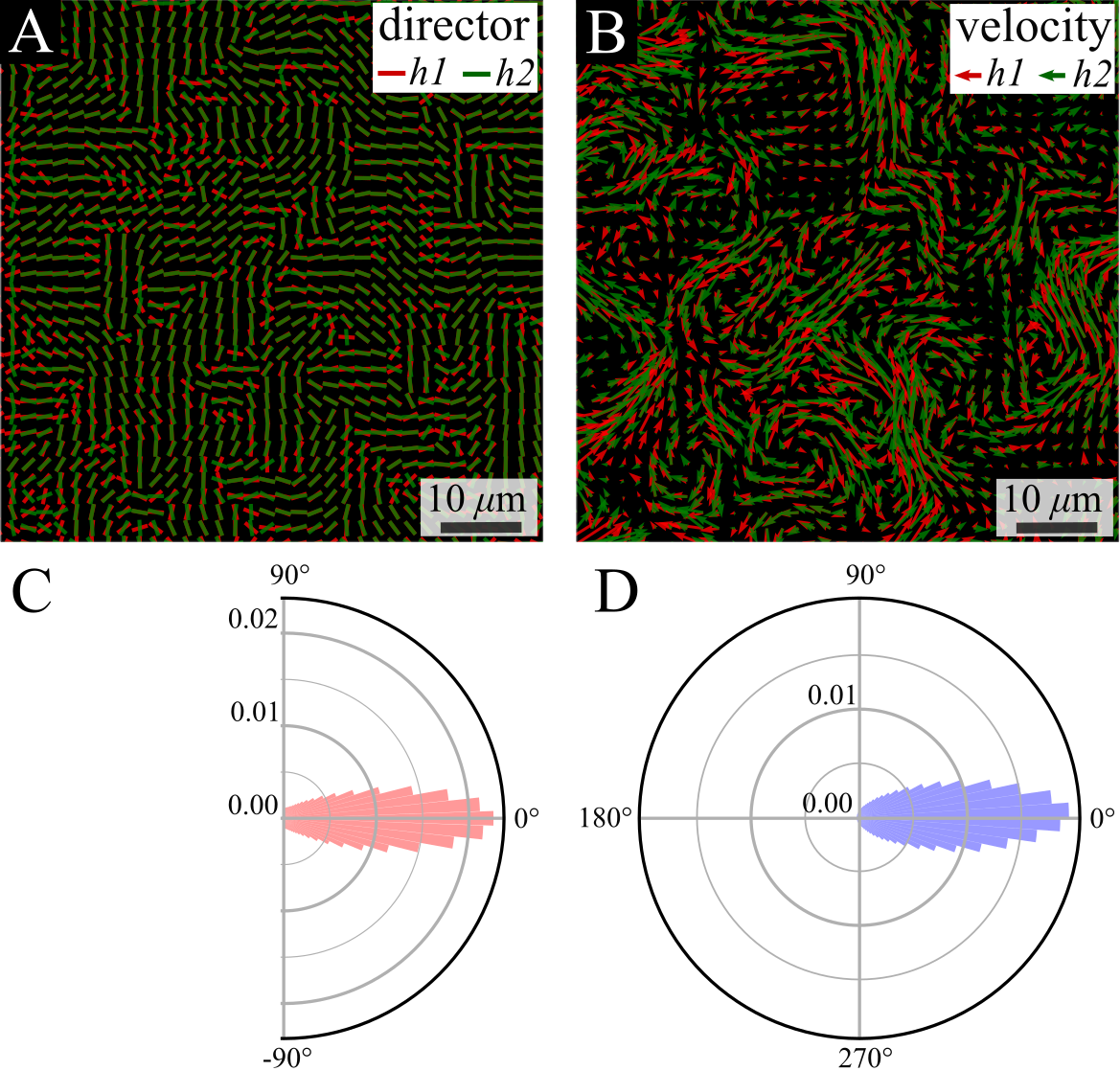}
  \caption{Comparison between layers---director and velocity fields. (A, B) Simultaneous snapshots showing the director and velocity fields at the bottom layer near the substrate ($h1$) and at the top layer ($h2$) of the swarm. (C, D) The probability density distribution of the difference in directions of the nematic director (C, red) and velocity (D, blue) fields between the bottom and top layers.
}
  \label{Fig.6}
\end{figure}

\begin{figure*}[h!]
\centering
  \includegraphics[width=0.75\textwidth]{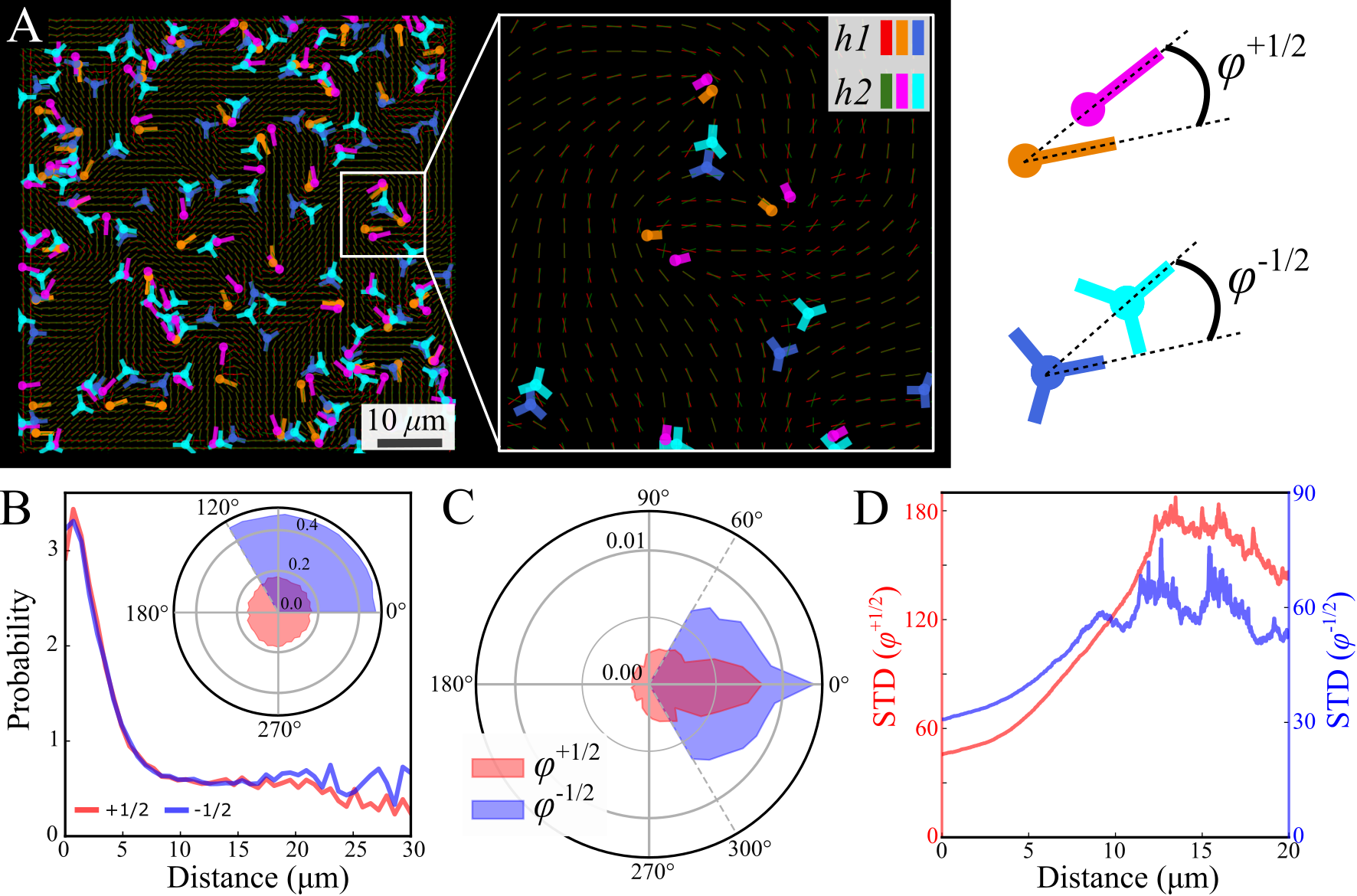}
  \caption{Comparison between layers---defects. (A) Simultaneous snapshots show defects at the bottom layer near the substrate ($h1$) and the top layer ($h2$). Inset A zoom-in illustrating top-bottom defect pairs of the same sign. The color code in the inset indicates to which layer the defects belong. (B) The displacement probability between pairs of defects ($+1/2$,$+1/2$ red and $-1/2$,$-1/2$ blue), is normalized by randomly distributed defects. A probability of 1 corresponds to an uncorrelated distribution. The inset shows a uniform probability density function (PDF) of displacement direction relative to the defect orientation. (C) PDF of the phase $\varphi$ between defect pairs ($\varphi^{+1/2}$ red and $\varphi^{-1/2}$ blue). The phase definition is illustrated in the sketch at the upper right. (D) Standard deviation of $\varphi$ as a function of their displacement ($\varphi^{+1/2}$ red and $\varphi^{-1/2}$ blue).
}
  \label{Fig.7}
\end{figure*}

The fit between theory and experiment is not complete. Figure \ref{Fig.4}A shows different MSD behavior for positive and negative defects, which we assume comes from the fact that active forces cause the $+1/2$ defects to self propel~\cite{pismen2013dynamics}. However, Giomi \cite{RefWorks:RefID:38-giomi2015geometry} demonstrates that on short time scales prior to annihilation, both defect types should appear super diffusive with similar power law exponents. This follows from the fact that the dynamics of the defects are also driven by elastic forces. These cause defects to continually move in a way that reduces the energy stored in the distortions of the director field, which would lead to super diffusive behaviour. A similar argument also applies to the orientations of defects, there are no net active torques on the defects and they continually rotate to minimize the elastic energy~\cite{RefWorks:RefID:46-pearce2021properties}. However, defects in our bacterial colonies appear to diffuse rotationally (Fig. \ref{Fig.4}B). The difference between theory and our experimental results may be evidence that the bacterial system has a small elastic coefficient and elastic interactions are screened over a short length scale leading to diffusive behaviour of defects in the bulk.

The experimental results suggest a new form of long-range interactions between defects, which is manifested in hyperuniformity (Fig.~\ref{Fig.5}). Hyperuniform structures are point patterns in which long-range spatial fluctuations are suppressed, as compared to an ideal gas \cite{RefWorks:RefID:57-torquato2018hyperuniform}. Formally, a system is said to be hyperuniform if its structure factor $S(q)$ vanishes for $q\xrightarrow ~0$. For charged particles, for example,
ionic solutions at equilibrium, one needs to consider the charge-density structure factor, $S_\rho(q)$. Within the Debye-Huckle theory, which is consistent at low concentrations, it can be computed analytically \cite{RefWorks:RefID:40-hansen2013theory}, giving rise to $S_\rho(q) \sim q^2$. Hence, at low concentrations ionic solutions are hyperuniform. It has been shown that the principle interaction between topological defects in active nematic systems is colomb-like \cite{RefWorks:RefID:35-decamp2015orientational,RefWorks:RefID:61-vromans2016orientational,RefWorks:RefID:46-pearce2021properties}. However, active nematic systems are far from equilibrium. Still, the charge-density structure factor obtained for topological defects in swarming bacteria is consistent with the result for the dilute colomb gas. Note that the lowest wave vector observed is bounded by the field of view. As a result, the reported experiment does not allow us to compute smaller wave vectors.

Lastly, we discuss some of the biological implications of our results. Recent experiments suggest that defects may serve a biological function. For example, it has been suggested that the centers of cell extrusion in epithelial monolayers \cite{RefWorks:RefID:50-saw2017topological,guillamat2022integer} act as local sources of cellular flows at the edge of fibrosarcoma cell colonies \cite{RefWorks:RefID:53-yashunsky2022chiral}. Defects were shown to shape the interface of bacterial colonies \cite{RefWorks:RefID:52-yaman2019emergence} and affect \textit{Hydra} morphogenesis \cite{RefWorks:RefID:67-maroudas-sacks2021topological}. In addition, defects were found to promote a transition from single-layer to multi-layer configuration in bacteria colonies  \cite{RefWorks:RefID:34-copenhagen2021topological} and in sheets of myoblast cells \cite{RefWorks:RefID:49-sarkar2023crisscross}. However, in this study, we found that $-1/2$ topological defects serve as sources for the coarse-graining bacterial flow (Fig.~\ref{Fig.3}F), in contrast to previous results for swimming bacteria, biofilms, and cell cultures which found a sink at $-1/2$ defects \cite{RefWorks:RefID:37-genkin2017topological, RefWorks:RefID:34-copenhagen2021topological, RefWorks:RefID:59-kawaguchi2017topological, RefWorks:RefID:53-yashunsky2022chiral}. 

For swarming bacteria, the biological function of topological defects is still unclear. We hypothesize that the main role of defects and their self-propulsion is in maintaining a nematic turbulent regime that leads to a well-mixed colony, which is important for the colony’s health, growth and survival. Furthermore, divergent flow at the core of $-1/2$ defect (Fig.~\ref{Fig.3}F) suggests they may also promote vertical mixing within a multilayered bacterial colony. We also observe indications that the topological defects are arranged in a hyperuniform manner (Fig.~\ref{Fig.5}), meaning this mixing effect is spread regularly across the colony. 

\section{Methods}
\subsection{Culture}
We work with \textit{B. subtilis} 3610, the “wild-type” (WT) strain, which is a rod-shaped flagellated species with dimensions \qtyproduct{1 x 7}{\um}. Cells are stored at \qty{-80}{\degreeCelsius} in frozen stocks. Fluorescently labeled variants, green and red, are used (amyE::PvegR0\_sfGFP\_spec and pAE1222-LacA-Pveg-R0\_mKate, MLS); labeling does not affect any known measured quantity. In some experiments, we mix the two labeled strains at ratio 1:1 (after separate overnight growth). Swarm experiments are typically done in a standard Petri-dish (\SI{8.8}{\cm} in diameter), where the colonies are grown on soft agar (0.5 \% and \SI{25}{\g/\l} LB) and aged in ambient lab conditions (\qty{24}{\degreeCelsius} and 35 \% RH) for 4 days. The latter forms the thicker colony structure compared to past results. Overnight cultures are grown from isolated colonies (which are grown from frozen stocks on hard agar plates (2 \% and LB)) at shaking (200 rpm and \qty{30}{\degreeCelsius}) for 18 h in LB liquid medium (\SI{2}{\ml} in a \SI{15}{\ml} tube). A small drop of overnight culture (\SI{4}{\ul}), either from a single strain, or a 1:1 mix of the two labeled variants, is deposited on the agar in the middle of the plate. The colonies grow for a few hours in a 95 \% RH incubator at \qty{30}{\degreeCelsius}. The swarm colonies form a quasi-2D structure with a thickness of $\sim$\SI{7}{\um}, which is uniform along centimeter-wide distances.

Observations are done using an optical microscope (Axio Imager Z2 Zeiss; $40\times$) operated at fluorescence mode; this magnification yields a \qtyproduct{150 x 150}{\um} observation frame. For the three-dimensional experiments, in the fluorescence mode, the system is equipped with a splitting device (Optosplit II) that enables dual excitation and acquisition. The two fluorescence fields are obtained for the same spatial field and time. The image is projected on a NEO camera with $1800\times 900$ pixels and at $50$ frames/sec so that the green cells are seen on the left panel while the red ones are seen on the right panel. The system operates with the dual fluorescence set Ex 59026x, beam splitter 69008bs, and Em 535/30; 632/60. A compensation lens (an integral part of the Optosplit II device) is used to adjust the focus in each of the panels so that the green panel shows only cells located at the bottom of the colony, and the red panel shows only cells located at the top of the colony, $\sim$\SI{7}{\um} above. Cells located $>$\SI{1}{\um} higher or lower from the observation plane are pale enough to not contribute to the analyses -- this was done by looking at a monolayer of cells deposited on the agar and traveling \SI{1}{\um} above or below yielding a vague image.

\subsection{Image Analysis} Image analysis was conducted using custom Python code. To obtain the velocity field from microscopic images, we employed the Lucas-Kanade optical flow method available from OpenCV library. The director field was computed with the second-moment matrix using the structure tensor function from the skimage Python library. Both flow and director fields were computed with an averaging window size of \qtyproduct{2.5 x 2.5}{\um}. \red{To estimate the errors in the detection algorithm we compare the number of defects in a single frame vs how many defects could be followed over at least 3 frames. The ratio is 0.942 for $+1/2$ defects and 0.939 for $-1/2$. Therefore, we estimate the total number of errors is less than 6 \%.}
\subsection{Defect detection and tracking} Nematic defects were found using the winding number method outlined in \cite{RefWorks:RefID:68-de1993physics}. \red{In essence, the winding number quantifies the degree of rotation exhibited by the nematic field around any closed loop of nearest neighbors. If that closed loop encircles a defect, the winding number will reflect the charge of that defect \cite{hoffmann2021robustness}}. The orientation of defects was calculated according to Vromans \& Giomi (2016) \cite{RefWorks:RefID:61-vromans2016orientational}. Defect trajectories were automatically identified by tracking defects using the TrackMate plugin in the ImageJ software \cite{RefWorks:RefID:60-ershov2022trackmate}.

Top–bottom defect pairs were identified by measuring the distance between defects in distinct layers of the bacteria colony. A pair is defined as the closest pair of defects with the same charge across the bottom and top layers. It is important to note that both layers were imaged simultaneously\red{, and each defect was paired to its same-charge closest neighbor (in the other layer). This simple algorithm introduces no bias in the relative orientations of the defects and has no cutoff length. Close to the point of annihilation or creation, it is (physically) possible that defects will not have a corresponding partner in the opposite layer of the nematic. This effect is expected to be very short lived due to the thin nature of the active nematic studied here, and is therefore not taken into account, see \cite{shankar2018defect} for details.}

\subsection{Average \red{director and} flow fields around defects} To measure the average fields around $\pm 1/2$ defects, we collected all the defects with the same topological charge and with their corresponding flow fields in a window of size \qtyproduct{35 x 35}{\um} centered at the defect core. \red{Then we rotate the local director and velocity fields to align all defects in the same direction. Finally, we calculate the average of the rotated fields.}
\subsection{Structure factor} The structure factor of $N$ particles with positions $\textbf{r}_i$ is defined as
\[ 
   S({\textbf q})=\langle \frac{1}{N}\bigg | \sum_{i=1}^{N} e^{-i {\textbf q} \cdot {\textbf r}_i} \bigg |^2 \rangle 
\]
where brackets denote ensemble averaging, here over frames in the experiments.
The cross-correlation structure factor in a binary mixture of $N$ particles with positions $\textbf{r}_i$ and $M$ particles with positions $\textbf{R}_i$ is
\[ 
   S_2({\textbf q})=\langle \frac{1}{N}\bigg | \sum_{i=1}^{N}\sum_{j=1}^{M} e^{-i {\textbf q} \cdot ({\textbf r}_i-{\textbf R}_j)} \bigg |^2 \rangle 
\]
Note that because $M$ and $N$ are practically the same, the cross-correlation structure factor of $+1/2$ defects around $-1/2$ is the same as for $-1/2$ defects around $+1/2$.

We denote by $S_{+1/2}(\textbf{q})$ the structure factor of the positions of $+1/2$ defects on their own,  by $S_{-1/2}(\textbf{q})$ the structure factor of the positions of $-1/2$ defects on their own, and by $S_{\pm 1/2}(\textbf{q})$ the structure factor of cross-correlations between $+1/2$ and $-1/2$ defects. In Fig. S1, we show results, average over wave vectors with a norm in the range $(q-q_{\rm min}/2,~q+q_{\rm min}/2]$, where $q_{\rm min}$ is the minimal wave vectors accessible by the observation frame, $q_{\rm min}=2\pi /L$, $L$ being the viewing length. Note that the zero mode, $q=0$, which is extensive, is not included. We denote by $S_{\rm both}(\textbf{q})$ the structure factor obtained for all defects, both $+1/2$ and $-1/2$ disregarding the signs. We denote by $S_\rho(\textbf{q})$ the structure factor for the charge density, defined as $S_\rho(\textbf{q})=S_{+1/2}(\textbf{q})S_{-1/2}(q)/2-S_{\pm 1/2}(\textbf{q})$.

\section*{Conflicts of interest}
There are no conflicts to declare.

\section*{Acknowledgments}
We thank Haim Diamant for discussions. The research was supported by The Israel Science Foundation (grants No. 838/23, No. 2044/23 and No. 832/23).






\bibliography{citations} 
\bibliographystyle{rsc} 

\end{document}